# Some Applications of Directed Signature Scheme

**Sunder Lal** [*] **and Manoj Kumar** [**]

[*] *Dept of Mathematics, IBS Khandari.Dr. B.R.A.University Agra.*
Sunder_lal2@rediffmail.com.in.
[**] *Dept of Mathematics, HCST, Farah – Mathura,* (U. P.) – 281122.
Yamu_balyan@yahoo.co.in.

**Abstract.** Directed signature scheme is applicable when the signed message contains information sensitive to the receiver, because only receiver can directly verify the signature and that he/she can prove its validity to any third party, whenever necessary.

This paper presents *two* applications of directed signature scheme. **(i)** *Directed –Delegated Signature Scheme*. This scheme combines the idea of proxy signatures with directed signature scheme. **(ii)** *Allocation of registration number.* This scheme proposes a registration scheme in which the registration number cannot be forged and misused.

## 1. Introduction

The most important part of a message is the signature of the sender. Usually written signature is hard to duplicate. Therefore this is a natural tool to **authenticate** the communication. Since physical signature is meaningless in electronic messages; one has to rely on other methods like digital signature.

Public key cryptography discovered by **W. Diffie** and **M. Hellman** [6] in 1976 has revolutionized the ways of message communications through insecure media. It is now possible for the people who have never met before to communicate with one another in a secure and authenticate way over an open and insecure network such as **Internet**. Thus there is a growing use of public key techniques in cryptographic applications. In particular, **digital signature** scheme using public key techniques is one of the most important cryptographic tools, which is essential in implementing various security measures and authentication.

Digital signature scheme allows a user with a **public key** and a corresponding **private key** to sign a document in such a way that everyone can verify the signature on the document (using her/his public key), but no one else can forge the signature on another document. This **self-authentication** is required for some applications of digital signatures such as certification, by some authority. In many situations, signed message is sensitive to the signature receiver. Signatures on medical records, tax information and most personal/business transactions are such situations. *Consider when a user* A *wants to generate a signature on a message m, sensitive for* B *and the message is also of concern to other users. For this situation, the form of the signature* *should be such that only* B *can directly verify the signature and that* B *can prove its validity to any third party* C, *whenever necessary. Such signatures are called* **directed signatures** [4, 5, 8, 9, 12]. In directed signature scheme, the signature receiver B has full control over the signature verification process. Nobody can check the validity of signature without his cooperation.

The concept of directed signatures was first presented by C.H.Lim and P.J. Lee [8]. It is a construction based on the GQ signature scheme [7]. D.Chaum [4] introduced the concept of designated confirmation. Later T. Okamoto [12] presented a more practical construction of designated confirmer signatures.

This paper presents two applications of directed signature scheme. The paper is organized as follows:-

The section-2 presents **some basic tools**. Section-3 describes a **Directed –Delegated Signature Scheme.** An application to **Allocation of registration number** is discussed in section-4**.**

## 2. Preliminaries

**2.1.** Throughout this paper we will use the following system setting.

- A prime modulous $p$, where $2^{511} < p < 2^{512}$;
- A prime modulous $q$, where $2^{159} < q < 2^{160}$ and $q$ is a divisor of $p - 1$;
- A number $g$, where $g \equiv k^{(p-1)/q} \mod p$, $k$ is random integer with $1 \leq k \leq p-1$ such that g >1; (g is a generator of order $q$ in $Zp^*$).
- A collision free one-way hash function $h$ [17];

The parameters $p, q, g$ and $h$ are common to all users. We assume that every user A chooses a random $x_A \in Zq$ and computes $y_A = g^{x_A} \mod p$. Here $x_A$ is the private key of A and $y_A$ is the public key of A. For our purpose, we use the directed signature scheme based on Schnorr's signature scheme [14]. These basic tools are briefly described below:-

### 2.2. Schnorr's signature scheme

In this scheme, the signature of A on message $m$ are given by ($r_A, S_A$), where,

$$r_A = h(g^{k_A} \mod p, m), \text{ and } S_A = k_A - x_A . r_A \mod p.$$

Here random $k_A \in Zq$ is private to A .The signature are verified by checking the equality

$$r_A = h(g^{S_A} y^{r_A} \mod p, m).$$

## 2.3. Directed signature scheme

Suppose that user A wants to generate a signature on message *m* so that only the receiver B can verify the signature and that B can prove the validity of signature to any third party C, whenever necessary. This scheme consists of the following steps.

### 2.3.1. Signature generation by A to B

(a). A picks at random $K_{a_1}$ and $K_{a_2} \in Zq$ and computes

$$W_B = g^{K_{a_1} - K_{a_2}} \mod p, \quad \text{and} \quad Z_B = y_B^{K_{a_1}} \mod p.$$

(b). Using a one-way hash function *h*, A computes $r_A = h(Z_B, W_B, m)$, and then

$$S_A = K_{a_2} - x_A \cdot r_A \mod q.$$

$\{S_A, W_B, r_A, m\}$ is the signature of A on *m*.

### 2.3.2. Signature verification by B

(a). B computes $\mu = [g^{S_A} (y_A)^{r_A} W_B] \mod p$, and $Z_B = \mu^{x_B} \mod p$.

(b) B computes $h(Z_B, W_B, m)$ and checks the validity of signature by equality

$$r_A = h(Z_B, W_B, m) \mod q.$$

### 2.3.3. Proof of validity by B to any third party C

(a) B sends to $\{S_A, W_B, r_A, m, \mu\}$ to C.

(b) C checks if $r_A = h(Z_B, W_B, m) \mod q$.

If this does not hold C stops the process; otherwise goes to the next steps.

(c) B in a zero knowledge fashion proves to C that $\log_\mu Z_B = \log_g y_B$ as follows:-

- C chooses random $u, v \in Zp$ computes $w = \mu^u \cdot g^v \mod p$, and sends *w* to B.
- B chooses random $\alpha \in Zp$ computes $\beta = w \cdot g^\alpha \mod p$, and $\gamma = \beta^{x_B} \mod p$, and sends β, γ to C.
- C sends *u, v* to B, by which B can verify that $w = \mu^u \cdot g^v \mod p$.
- B sends α to C, by which she can verify that

$$\beta = \mu^u \cdot g^{v+\alpha} \mod p, \quad \text{and} \quad \gamma = Z_B^u \, y_C^{v+\alpha} \mod p.$$

## 3. Directed Delegated Signature Scheme

This section combines the idea of proxy signature scheme [10, 16] with directed signature scheme and obtains a **Directed-delegated signature scheme.** In this application of directed signature scheme, the signing of message is done in two phases. The first phase is off-line. This phase is governed by proxy signature scheme, and can be performed even before the message to be signed is known. The second phase is on-line. It starts after the message becomes known and utilizes the precomputation of the first phase.

Delegation of rights is a common practice in the real world. In the electronic world to facilitate this requirement, proxy signature scheme have been invented to delegate signing capability efficiently and transparently. Digital signature schemes rely on a secret signature key [1], which only the certified person knows. If this secret key is delegated to another person directly, it can no longer be identified with that person and hence the assumption of the digital signature is broken. We therefore need to delegate the signing capability without revealing the secret key such that the recipient can verify the signature of the original signer with the help of proxy signer.

This section proposes an application of *directed-delegated signature scheme*. In this scheme, CMO is a central authority, who designates a trusted authority B to issue certificate on his behalf. The contents of the certificate are message, which is sensitive for a patient C and is also of concern to other users. C wants a certificate that nobody can check the validity of signature without his help, but C can prove the validity of the signature to any third party Yamu, whenever necessary.

Consider a situation, there is a NGO'*s* hospital which facilitate AIDS checkup. The hospital is headed by a CMO A, who has many responsibilities to perform. He designates a trusted colleague B to issue certificate on his behalf. A patient C does not want to disclose the result of his/her check-up and he/she want a certificate that nobody can read without his help. Unfortunately if C is HIV positive then there is need of curing. For treatment there is another important hospital headed by a NGO chief Yamu, Y, Where all types of Medicine, remedies and all resources are available for AIDS patients C. For treatment there, C has to prove the validity of a certificate to a desired person (Yamu) only.

The one solution of such problems is governed by directed signature schemes. Here the receiver C can directly verify the signature and that C can prove its validity to any third party Yamu, whenever necessary. Since directed signature can be verified with the help of receiver C only so we can say that the contents of certificates have no validity without the signature verification.

The following is an exposition on how proxy and directed signature scheme can be implemented for our construction.

## 3.1 Application of directed-delegated signature scheme

Before organizing the check up, A appoints B as a designated signer. He delivers a designated signature key to B using the following protocol.

### 3.1.1. Signature key delegation by A

1. A selects $k_A \in Zq$ computes $r_A = g^{k_A} \mod p$ and sends $r_A$ to B.

2. (a). B randomly selects $\alpha \in Zq$ and computes $r = g^{\alpha} r_A \mod p$.

   (b) If $r \in Zq^*$, he sends $r$ to A, otherwise goes to step (a)

3. A computes $s_A = r x_A + k_A \mod q$ and forwards $s_A$ to B.

4. B computes $S = s_A + \alpha \mod q$ and check if $g S = y^r r \mod p$

If equality holds, B accepts '$s_A$' as a valid designated signature key from A. Now the hospital is open for the public. B does the check up of a patient C and make the message $m$. B gives a certificate with message $m$ to C. The signature generation and verification is done by the following protocol.

### 3.1.2. Signature generation by B for C

(a) B picks at random $K_{b_1}$ and $K_{b_2} \in Zq$ and computes

$$W_B = g^{K_{b_1} - K_{b_2}} \mod p \text{ and } Zc = y_C^{K_{b_1}} \mod p.$$

(b) B computes $r_B = h(Zc, W_B, m)$ and $S_B = K_{b_2} - S \cdot r_B \mod q$.

(c) B sends to C $\{ S_B, W_B, r_B, m \}$ as CMO signature.

### 3.1.3. Signature verification by C

(a) C computes $\mu = (g^{S_B} (y_A^r \cdot r)^{r_B} W_B) \mod p$ and $Zc = \mu^{x_C} \mod p$.

(b) C checks that $r_B = h(Zc, W_B, m)$, for the validity of signature.

### 3.1.4. Proof of validity by C to Y

This is done using the protocol discussed in the section- 2.3.3.

## 3.2. Security discussions

This signature scheme is secure if existential forgery (providing a new message –signature pair) is computationally infeasible. In this section, we discuss some possible attacks.

(a). If the designated proxy signer B is dishonest then he can cheat the original signer A and get her/his signature ($r$, $s_A$) on any message $m$ of her/his choice.

The solution of this problem is the existence of a trusted third party. The original signer A may stress that all messages between two parties A and B during the key delegation protocol be authenticated. The third party keeps the records of original signer's orders, and checks any case of designated signer disobeying original signer's order.

(b). Can one get integer $K_{b_2}$ and $S$ (secret signature key of proxy signer), from the equation

$$S_B = K_{b_2} - S \cdot r_B \mod q \ ?$$

Here the numbers of unknown parameters are two. The number of equation is one, so it is computationally infeasible for a forger to collect the secret integer $K_{b_2}$ and $S$.

(c). Can one impersonate the designated signer B ?

A forger may try to impersonate the designated signer B by randomly selecting two integers $K_{i_1}$ and $K_{i_2} \in Z_q$. But without knowing the secret part $\alpha$, it is difficult to generate a valid proxy signature key $S$ and $S_B$ to satisfy the verification equation,

$$Z_C = (g^{S_B} (y_A^{\ r} \cdot r)^{r_B} W_B)^{x_C} \mod p, \quad r_B = h(Z_C, W_B, m).$$

(d). Can one forge a signature { $S_B$, $W_B$, $r_B$, $m$, $r$} using the equation,

$$\mu = (g^{S_B} (y_A^{\ r} \cdot r)^{r_B} W_B) \mod p \ ?$$

To compute the integer $S_B$ from this equation is equivalent to solving the discrete logarithm problem. If any forger randomly selects $S^*$ and sends { $S^*$, $W_B$, $r_B$, $m$, $r$ } to B, the receiver B computes

$$\mu^* = [g^{S^*} (y_G)^R W] \mod p, \quad Z^* = \mu^{* \, x_B} \mod p.$$

and can check if $r_B = h(Z^*, W_B, m)$, to detect the forgery.

### 3.3. Illustration

We illustrate the above scheme using small parameters. Taking $p = 23$, $q = 11$, $g = 6$ and the secret keys and the public keys of users are as follows.

| Users | Secret key | public key |
|-------|------------|------------|
| A | 3 | 9 |
| B | 5 | 2 |
| C | 6 | 12 |
| Y | 8 | 18 |

#### 3.3.1. B is appointed as a proxy signer by A

1. A selects $k_A = 7$ and sends $r_A = 3$ to B.

2. B selects $\alpha = 5$ and sends $r = 6$ to A.

3. A computes $s_A = 3$ and sends to B.

4. B computes proxy signature key $s = 8$ and checks if $6^8 = [(9^6 . 6)]$ Mod23.

#### 3.3.2. Signature generation by B for C

(a) B picks at random $K_{b_1} = 7$, $K_{b_2} = 2$ and calculate $W_B = 2$ and $Z_C = 16$.

(b) B computes $r_B = h(16, 2, 0, 8, 3, 18) = 2$.

(c) B computes $S_B = 8$, and sends {2,2,8,0,8,3,18,6} to C.

#### 3.3.3. Signature verification by C

(a) C computes $\mu = 3$ and $Z_C = 16$.

(b) C checks $r_B = h(16, 2, 0, 8, 3, 18) = 2$, for the validity of signature.

#### 3.3.4. Proof of validity by C to Y

(a) C sends (16,2,2,8,3,0,8,3,18,3) to Y.

(b) Y checks $r_B = h(16, 2, 0, 8, 3, 18) = 2$,

   If this does not hold stops the process; otherwise goes to next step.

(c) Now C can prove that $\log_3 16 = \log_6 12$, in a zero knowledge fashion by using the following confirmation protocol.

(i). Y chooses at random $u = 13, v = 15$ and computes $w = 3$ and sends $w$ to C.

(ii). C chooses at random $α = 8$, computes $β = 8$ and $γ = 13$ and sends $β, γ$ to Y.

(iii). Y sends $u, v$ to C, by which C can verify that $w = 3$.

(iv). C sends $α$ to Y, by which she can verify that $β = 16$ and $γ = 4$.

## 3.4. Remarks

In this section, we have discussed an on-line /off-line directed signature scheme, which is useful in that case when the signed message is sensitive to signature receiver. In this scheme, the signature receiver C has full control over the signature verification process. Nobody can check the validity of signature without his co-operation. The receiver is also able to prove the validity of the signature, whenever necessary.

We have presented a construction of such a directed signature scheme, which is based on discrete logarithm problems. Hence the security level of this scheme is similar to that of other scheme based on discrete logarithm. Since the relation between the signer and signer's secret key is not known to anyone, this scheme is more secure than any other scheme, based on the discrete logarithm.

## 4. Allocation of registration number

Registrations of various kinds are a common practice in our society, like that of vehicle, shop and factory etc. In daily life, there are so many situations, when it is necessary, beneficial and expedient to have a registration number for vehicles etc.

This section proposes a registration scheme in which the registration number cannot be forged and misused. Under this scheme the validity of an allocated registration number can be verified at any time by any authority. The allocating authority and verifying authority may be different. For the practical implementation of this idea, we use a directed signature scheme.

We all are familiar with the present status of our registration system. A hand written signature is used for the allocation of registration number by the authority. Every signature is followed a lot of formalities and records. Unfortunately the present system is not much secure and is liable.

We assume a government center, providing the registration number for the public. An officer Yamu, Y, heads this center. Y possesses a secret and public key pair $(x_o, y_o)$.

Again consider a public person Chaya, C, with a secret and public key pair $(x_c, y_c)$ wants her registration number. The officer, Y generates a registration number with message $m$, so that C can directly collect her registration number. She can use her registration number publicly. She is able to prove its validity to any authorized party R whenever necessary. No one other than C can use this registration number because only she can prove its validity. This section is organized as follows: –

Allocation of registration number, and verifying processes are as follows.

### 4.1.1. Allocation of registration number by Y to C

(a). Y picks at random $K_{y_1}$ and $K_{y_2} \in Zq$ and computes

$$Wy = g^{K_{y_1} \cdot K_{y_2}} \mod p \quad \text{and} \quad Zc = y_c^{K_{y_1}} \mod p.$$

(b). Y again computes $r_y = h(Zc, Wy, m)$ and $S_y = K_{y_2} - x_o \cdot r_y \mod q$.

(c). Y sends $\{S_y, W_y, r_y, m,\}$ to C as her registration number.

### 4.1.2. Collecting and verification of registration number by C

(a). C collects $\{S_y, W_y, r_y, m\}$ and make this public as her registration number.

(b). C computes $\mu = [g^{S_y}(y_0^{r_y})W_y] \mod p$, $Zc = \mu^{x_c} \mod p$ and checks the validity of her registration by computing $r_y = h(Zc, Wy, m)$.

### 4.1.3. Verification Of registration number by authority R
This part is same in the previous section- 2.3.3.

### 4.2. Illustration

The following illustration supports our scheme for practical implementation. Taking $p = 23$, $q = 11$ and $g = 5$. The secret and private key of users is as follow

| For | Secret key | private key |
|-----|------------|-------------|
| Y   | 5          | 20          |
| C   | 8          | 16          |

### 4.2.1. Allocation of registration number by Y to C

(a) Y picks at random $K_{y_1} = 7$, $K_{y_2} = 4$ and calculate

$$W_y = 10, \quad Z_y = 1 \quad \text{and} \quad r_y = 2. \text{ (taking } m = 1\text{)}$$

(b) Y computes $S_y = 5$, and sends $\{10,2,5,1\}$ to C as her registration number.

### 4.2.2. Verification of registration number by C

(a). C collects her registration number {10,2,5,1} and makes this public.

(b). C checks the validity of her registration by computing $Z_C = 18$ and check if $r_y = 2$.

### 4.2.3. Validity proof of registration number by C to any authorized party R

(a) C computes $\mu = 6$, and $Z_C = 6^8 \mod 23 = 18$. and sends (18,10,2,5,1) to R.

(b) R checks if $r_y = h(18,10,1) = 2$.

   If this does not hold stops the process; otherwise goes to next step.

(c) Now C proves to 'R' that $\log_6 18 = \log_5 16$, in a zero knowledge fashion by using the confirmation protocol. The illustration of this protocol is similar to that of previous in section- 3.3.4.

## 4.3. Remarks

Thus above construction facilities the allocation of registration number in the electronic world with the following characteristics.

- Only the user can use his/her registration number, due to the property of directed signature scheme.

- The problems of forgery can be solved easily.

- By using this scheme, we can minimize the possible misuse of the present system.

- The obvious advantage of our scheme over present system is that the resulting registration number has no meaning to any third person.

- Since the relation between the signature and the signer secret key is not known to anyone but the designated receiver. Hence security level is much higher than any other scheme based on discrete logarithm.

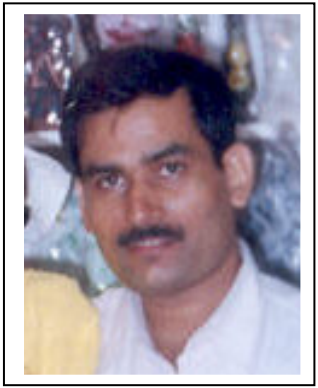

**Manoj Kumar** received the B.Sc. degree in mathematics from Meerut University Meerut, in 1993; the M. Sc. in Mathematics (Goldmedalist) from C.C.S.University Meerut, in 1995; the M.Phil. (Goldmedalist) in *Cryptography*, from Dr. B.R.A. University Agra, in 1996; submitted the Ph.D. thesis in *Cryptography*, in 2003. He also taught applied Mathematics at DAV College, Muzaffarnagar, India from Sep, 1999 to March, 2001; at S.D. College of Engineering & Technology, Muzaffarnagar, and U.P., India from March, 2001 to Nov, 2001; at Hindustan College of Science & Technology, Farah, Mathura, continue since Nov, 2001. He also qualified the *National Eligibility Test* (NET), conducted by *Council of Scientific and Industrial Research* (CSIR), New Delhi- India, in 2000. He is a member of Indian Mathematical Society, Indian Society of Mathematics and Mathematical Science, Ramanujan Mathematical society, and Cryptography Research Society of India. His current research interests include Cryptography, Numerical analysis, Pure and Applied Mathematics.